\documentclass[10pt,conference]{IEEEtran}
\IEEEoverridecommandlockouts
\usepackage{cite}
\usepackage{amsmath,amssymb,amsfonts}
\usepackage{algorithm}
\usepackage{algorithmicx}
\usepackage{algpseudocode}
\usepackage{graphicx}
\usepackage{booktabs}
\usepackage{textcomp}
\usepackage{xcolor}
\usepackage{hyperref}
\usepackage{comment}
\usepackage[nolist]{acronym}
\usepackage{paralist}
\usepackage[skip=0.333\baselineskip]{caption}
\usepackage{subcaption}
\usepackage{wrapfig}
\usepackage{placeins}
\def\BibTeX{{\rm B\kern-.05em{\sc i\kern-.025em b}\kern-.08em
    T\kern-.1667em\lower.7ex\hbox{E}\kern-.125emX}}

\begin{document}

\acrodef{DSO}{Distribution System operator}
\acrodef{DBSCAN}{Density-Based Spatial Clustering of Applications with Noise}
\acrodef{HDBSCAN}{Hierarchical Density-Based Spatial Clustering of Applications with Noise}
\acrodef{TCP}{Transmission Control Protocol}
\acrodef{TTL}{Time to Live}
\acrodef{IQR}{Interquartile Range}
\acrodef{MTU}{Master Terminal Unit}
\acrodef{VRTU}{Virtual Remote Terminal Unit}
\acrodef{RTU}{Remote Terminal Unit}
\acrodef{DBCV}{Density-Based Clustering Validation}
\acrodef{DoS}{Denial of Service}
\acrodef{SSL}{Secure Sockets Layer}
\acrodef{IEC104}{IEC 60870-5-104}
\acrodef{TLS}{Transport Layer Security}
\acrodef{SCADA}{Supervisory Control and Data Acquisition}
\acrodef{DER}{Distributed Energy Resources}
\acrodef{ICT}{Information and Communication Technologies}
\acrodef{CIA}{Confidentiality, Integrity and Availability}
\acrodef{ICS}{Industrial Control System}
\acrodef{IT}{Information Technology}
\acrodef{OT}{Operational Technology}
\acrodef{MulVAL}{Multi-host, Multi-stage Vulnerability Analysis}
\acrodef{NVD}{National Vulnerability Database}
\acrodef{OVAL}{Open Vulnerability and Assessment Language}
\acrodef{TTC}{Time-to-Compromise}
\acrodef{IDS}{Intrusion Detection System}
\acrodef{CVE}{Common Vulnerabilities and Exposures}
\acrodef{CVSS}{Common Vulnerability Scoring System}
\acrodef{Ac}{Access Complexity}
\acrodef{Au}{Authentication}
\acrodef{Ex}{Exploitability}
\acrodef{CPT}{Cyber-Physical Digital Twin}
\acrodef{DT}{Decision Tree}
\acrodef{GAN}{Generative Adversarial Network}
\acrodef{RF}{Random Forest}
\acrodef{SVM}{Support Vector Machine}
\acrodef{MCC}{Matthews correlation coefficient}
\acrodef{AUC}{Area Under Curve}
\acrodef{ROC}{Receiver Operating Characteristic}
\acrodef{TP}{True Positive}
\acrodef{TN}{True Negative}
\acrodef{FP}{False Positive}
\acrodef{FN}{False Negative}
\acrodef{CNB}{Complement Naïve Bayes}
\acrodef{XGB}{Extreme Gradient Boosting}
\acrodef{MQTT}{Message Queuing Telemetry Transport}
\acrodef{TTC}{Time-to-Compromise}
\acrodef{ML}{Machine Learning}
\acrodef{HMI}{Human-Machine-Interface}
\acrodef{CI}{Confidence Interval}

\acrodef{LLM}{Large Language Model}
\acrodef{DMZ}{Demilitarized Zone}
\acrodef{DoS}{Denial of Service}
\acrodef{DDoS}{Distributed Denial of Service}
\acrodef{CIA}{Confidentiality-Integrity-Availability}
\acrodef{PV}{Photovoltaic}
\acrodef{VM}{Virtual Machine}
\acrodef{OS}{Operating System}
\acrodef{RAG}{Retrieval Augmented Retrieval}

\acrodef{SSH}{Secure Shell}
\acrodef{LLC}{Logic Link Control}
\acrodef{ARP}{Address Resolution Protocol}

\bstctlcite{IEEEexample:BSTcontrol}

\title{
AI-based Attacker Models for Enhancing Multi-Stage Cyberattack Simulations in Smart Grids Using Co-Simulation Environments
}

\author{
\IEEEauthorblockN{%
Ömer Sen\IEEEauthorrefmark{1}\IEEEauthorrefmark{2},
Christoph Pohl\IEEEauthorrefmark{1},
Immanuel Hacker\IEEEauthorrefmark{1}\IEEEauthorrefmark{2},
Markus Stroot\IEEEauthorrefmark{1}\IEEEauthorrefmark{2},
Andreas Ulbig\IEEEauthorrefmark{1}\IEEEauthorrefmark{2},
}

\IEEEauthorblockA{%
\IEEEauthorrefmark{1}\textit{RWTH Aachen University,} Aachen, Germany |
\IEEEauthorrefmark{2}\textit{Fraunhofer FIT,} Aachen, Germany\\
Email: \{oemer.sen, immanuel.hacker, markus.stroot, andreas.ulbig\}@fit.fraunhofer.de} christoph.pohl@rwth-aachen.de
}

\maketitle

\begin{abstract}
The transition to smart grids has increased the vulnerability of electrical power systems to advanced cyber threats. To safeguard these systems, comprehensive security measures—including preventive, detective, and reactive strategies—are necessary. As part of the critical infrastructure, securing these systems is a major research focus, particularly against cyberattacks. Many methods are developed to detect anomalies and intrusions and assess the damage potential of attacks. However, these methods require large amounts of data, which are often limited or private due to security concerns. We propose a co-simulation framework that employs an autonomous agent to execute modular cyberattacks within a configurable environment, enabling reproducible and adaptable data generation. The impact of virtual attacks is compared to those in a physical lab targeting real smart grids. We also investigate the use of large language models for automating attack generation, though current models on consumer hardware are unreliable. Our approach offers a flexible, versatile source for data generation, aiding in faster prototyping and reducing development resources and time.
\end{abstract}


\section{Introduction}\label{sec:intro}

As power grids increasingly incorporate \ac{ICT}, they become more interconnected with external devices, removing the natural barrier of isolation and exposing them to new cybersecurity threats. A notable example is the 2015 Ukraine cyberattack, which resulted in a significant blackout. Cybersecurity in power systems is a vital, ongoing process based on prevention, detection, and reaction, guided by standards such as IEC 62351 and IEC 62443. Additionally, improving cybersecurity resilience and protecting critical infrastructures are emphasized legally with emerging regulations like NIS2.

Smart grids are complex systems and combine a lot of software and components.
They include their accumulated vulnerabilities and even the used protocols like IEC 104, commonly used in SCADA systems, lack authentication and encryption.
Implementing security enhancements means evaluating new metrics, analyzing component behavior in a network, examining logs, and running \acp{IDS} in smart grids.
These steps are promising but all depend on data or reference values for analysis. While some openly accessible datasets exist, many are either poorly maintained, too specific, or have other shortcomings \cite{kenyon2020public}.

Grid operators tend to keep their data private due to security concerns, forcing researchers to either build their own laboratories or simulate smart grids themselves. Modelling smart grids in a laboratory requires expertise, time to install hardware, and significant expense to operate. Implementing and setting up even a small laboratory demands professionals from various domains.


Other studies evaluated multiple simulation methods with and without cyberattacks \cite{le2019smart}. Some tools could generate reports and run simple \ac{DDoS} attacks, while others focused solely on co-simulating power grids and their network communication.
Manual red teaming approaches face challenges in consistency and scalability, making automation a promising solution. Frameworks like Plan2Defend \cite{plan2defend} and autonomous agents for cyber defense \cite{vyas2023automated} use autonomous methods to mitigate ongoing cyberattacks or execute plans, reducing response time. While these methods evaluate system security and monitor activity, they lack the ability to generate attack data, relying on manual implementation of sophisticated attacks for training.

Proactive testing via attack emulation allows for hardening security before actual attacks occur. Although individual cyberattacks on smart grids have been analyzed, there is less focus on general frameworks for complex data generation. \acp{IDS} require extensive data to evaluate performance and adapt to varying network topologies.


To address these challenges, we propose a method for automatically running cyberattacks inside a simulation environment, generating network logs and power simulation results. This approach allows for the creation of datasets in configurable network topologies and various attack scenarios, improving data availability and eliminating the need for expensive setups.
This comprehensive approach enables the extraction of valuable data, facilitating the study and enhancement of smart grid cybersecurity. Our contributions are the following:

\begin{itemize}
    \item Definition of the problem and corresponding requirements for our method, and deduction of a general concept for our work.
    \item Design and assessment of the AI-based attacker model in a simulation environment, detailing the physical laboratory and co-simulation structure, the autonomous agent's operation, deployment environments, and the data generation process.
    \item Evaluation of the data generation process via comparative analysis with a physical laboratory using a digital twin approach.
\end{itemize}

\section{Background}\label{sec:background}
\subsection{Smart Grid Control Architecture} \label{subsec:background_smartgrid}
Compared to early power grids that directly connected energy generators with consumers, modern energy infrastructure is much more complex. The increasing use of energy sources such as \ac{PV} systems and wind turbines makes regulation and coordination more challenging. When these generators are distributed and connected to a large power grid without bundling their loads in a large converter, special regulation and balancing are required before their energy can be fed into the grid. Smart grids provide a solution as cyber-physical systems that connect energy sources with consumers and enable communication between actors to balance and control the power grid reliably.


Understanding smart grid operation requires knowledge of its components:

\textbf{Terminal Units} enable smart grid communication by collecting and transmitting data and sending and receiving control commands. These units are divided into \acp{RTU}, connecting to physical equipment, and \acp{MTU}, aggregating data from \acp{RTU}.
\textbf{Supervisory Control and Data Acquisition (SCADA) Systems} handle data from terminal units, process it to adjust grid behavior, and provide \acp{HMI} for data visualization and remote control of grid components.
\textbf{Intelligent Electronic Devices (IEDs)} are controllers in power equipment that collect data from sensors and issue commands. They communicate with SCADA systems directly or via \acp{RTU}.
\textbf{Decentralized Energy Resources (DERs)} are small components supplying or demanding electricity, connecting directly to the grid or aggregated. Examples include microturbines, solar arrays, and energy storage systems.
\textbf{Protocols} are crucial for data exchange in smart grids. Common ones like IEC 104 and Modbus operate over TCP/IP networks to connect SCADA systems with terminal units but lack integrated security, making them vulnerable to cyberattacks. Secure Shell (SSH) is used for remote access but can pose security threats if compromised.

The overall structure of an Industrial Control System (ICS) can be described using the Purdue model, proposed by Williams et al. in the 1990s \cite{williams1994purdue}.  This model segments the network into multiple layers, allowing separation between operational technology (OT) and information technology (IT) systems. The model defines a hierarchy for communication between layers, ranging from external services at the highest level (Level 5) to the actual control of physical actors at the lowest level (Level 0).

The layers of the Purdue model are as follows:

\textbf{Level 5 (Enterprise Zone)} includes general IT systems accessible to external clients, providing the most open services. 
\textbf{Level 4} focuses on data and software for production scheduling and operational management, including mail servers and other widely accessible services.
The \textbf{Demilitarized Zone (DMZ)} separates IT and OT systems, using firewalls to block most communication protocols and secure sensitive data by preventing outgoing communication.
\textbf{Level 3 (Manufacturing Zone)} is the highest level within the OT site, containing local workstations and data collections for process data.
\textbf{Level 2} houses area control systems that collect and monitor data within subsystems.
\textbf{Level 1} consists of controlling elements that send measured values to data collectors and control physical actors and sensors.
\textbf{Level 0} includes all machinery performing physical tasks and measurements.

Despite being proposed decades ago, the Purdue model remains widely applied, particularly in the context of cybersecurity for cyber-physical systems.

\subsection{Simulation Aspect} \label{subsec:background_simulation}

To simulate smart grids, we must understand what simulation entails and how it can be achieved. Generally, simulation involves imitating real-world systems via software. Instead of creating a single, complex model, implementing smaller components separately and combining them to model complex systems increases the framework's sustainability and performance feasibility. These components can run in individual virtual environments to prevent interference.

Virtualization involves running individual processes or entire \acp{OS} on multiple virtual computers, each using a portion of the host computer's resources. 
In contrast to the well-established approach of bundling and isolating software in \acp{VM}, we implemented our work with the help of containerization. While VMs include their own operating systems, containerization allows software to share resources with the host dynamically, increasing performance for parallel execution. These instances are called containers and are built according to a blueprint, that defines the corresponding software, memory and network available when instantiating it. Containers can communicate through virtual network, sending actual network packets between specific instances.


Virtual networks allow communication between specific containers or VMs, creating separated networks on the same host. Orchestration tools further streamline the setup by reading configuration files to define containers and their settings, automating their deployment and management. 

Using container orchestration, we can automate the deployment of pre-configured containers and their virtual networks. This enables running complex scenarios with shared computing resources, making large-scale simulations less resource-demanding.

A key goal of our work is to run simulations automatically and incorporate systems classified as artificial intelligence (AI). 
\textbf{Automation} in software refers to the automatic operation of software with minimal user interaction, where a program executes pre-defined tasks.
\textbf{Autonomy} describes a program's ability to make decisions and react to its environment. According to \cite{formosa2021robot}, autonomy is a spectrum, with greater consideration of parameters leading to higher autonomy.

The European Commission's Artificial Intelligence Act \cite{EU-AIA} defines AI systems as machine-based systems designed to operate with varying levels of autonomy. These systems may exhibit adaptiveness and generate outputs such as predictions, content, recommendations, or decisions that influence environments.

\subsection{Multi-Staged Cyberattacks}

As smart grids increasingly resemble traditional computer networks, they become susceptible to similar cyberattacks \cite{li2012}. Below are some prevalent attack schemes and their potential combinations for more complex routines.

    
    
    
    
    

Attackers often combine techniques from different categories to execute complex attacks, making them harder to detect and defend against.

Advanced Persistent Threats (APTs) involve multiple stages executed sequentially. These stages typically include reconnaissance, initial access, persistence, and command-and-control, followed by the final impact. The complexity of APTs, with their numerous specialized steps, targets specific victims over a prolonged period, making them difficult to track and defend against.

For instance, the DARPA 2000 dataset features various \ac{DDoS} attacks comprising multiple steps \cite{ahmadian2016causal}. A simpler attack might include IP address scanning, service exploitation, and a subsequent \acp{DDoS} attack using compromised hosts. A more sophisticated attack, such as island-hopping, spreads malware through infected PDFs sent across the network until a desired database service is found. These examples highlight how combining different attack methods and exploiting vulnerabilities can allow attackers to gather extensive information and manipulate network components.

When trying to breach complex systems, such as critical infrastructure, attacks need a methodical progression.
So called \textit{multi-staged cyberattacks} unfold in several distinct steps, where each has an individual objective.
The key property of running complex cyberattacks in this manner is the dependency of consecutive steps on the results of their predecessors.

\begin{table}
    \centering
    \caption{Example for a multi-staged cyberattack.}
    \begin{tabular}{rlll}
            & Step                      & Result                & Depends on\\
            \hline\\
        1. & Get hostname               & Hostname              & \\
        2. & Check interfaces           & Available networks    & (1)\\
        3. & Scan network               & Reachable targets     & (2)\\
        4. & Generate SSH keys          & SSH keys              & \\
        5. & Brute-Force credentials    & Compromised targets   & (3),(4) \\
        6. & Connect agent to C2 server &                       & (5) \\
        7. & Remove SSH key             &                       & \\
    \end{tabular}
    \label{tab:multistaged-attack}
\vspace{-1em}
\end{table}

Such dependencies allow individual stages of the attack to rely on information that were previously collected and thus make modelling complex attacks possible.
This comes at the cost of having to plan the steps according to the available information and which dependencies are fulfilled.
In \autoref{tab:multistaged-attack} we picture a multi-staged cyberattack, where for example the brute-force depends on the results of a network scan. Running a brute-force attack without having a target diminishes the value of the operation; therefore, we can only perform this step if a potential target has been found.
Structuring attacks in such a modular way qualifies multi-staged cyberattacks to adapt the environment if planned accordingly.

\section{Methodology}




\subsection{Requirements}

We aim to create a framework for generating network logs of smart grids during multi-stage cyberattacks. In order to achieving this, there are a few requirements that we postulate from sparsely available datasets.

\textbf{Virtual Smart Grid Simulation:} A simulation resembling real grids, incorporating essential components of real-world smart grids, and entirely virtual to eliminate hardware requirements is necessary.
\textbf{Recordable Network Communication:} The simulation must log network traffic at the transport layer.
\textbf{Integrated Attack Simulation:} Capturing anomalies during adversary actions within the simulation is crucial for utilizing the network and component logic effectively.
\textbf{Realistic Impact on Data Output:} Generated data should be as usable as real hardware data, accurately reflecting cyberattack impacts.
\textbf{Configuration-Based Setup:} The simulation should be configurable for various scenarios and topologies, allowing automatic setup based on user-defined parameters to ensure scalability and prevent initialization errors.
\textbf{Autonomous Attacks with Modular Methods:} Supporting multiple attack methods, the simulation should manage multi-stage attacks autonomously, ensuring reproducibility and optimal method selection.
\textbf{Offline Availability:} The simulation should function without internet access, adhering to the segmented network architecture typical of smart grids.
\textbf{Clean and Consistent Network Logs:} Generated network traffic should be uniform, minimizing fluctuations across different runs to ensure attack impacts are clear.

For our data, two main properties are crucial: realism and consistency across different scenarios. Realistic data must closely depict the individual nuances of each attack, and consistency ensures that data output is structured similarly across different runs. We identified four main metrics to analyze each attack scenario:

\textbf{Power Simulation Results:} The simulation must accurately reflect the effects on power generation and distribution, showing changes in component behavior due to cyberattacks.
\textbf{Timely Spread:} Attacks should occur within a consistent timeframe, with tasks performed at similar points during the simulation.
\textbf{Traffic Volume:} Reliable simulation behavior is indicated by consistent data communication volume and direction during attacks.
\textbf{Protocol Distribution:} Consistent protocol usage during an attack ensures stable behavior of the involved components.

These metrics help ensure that the generated data is both realistic and consistent, providing a robust foundation for analyzing and improving smart grid cybersecurity. Our methodology focuses on these requirements, and we will evaluate how these were satisfied by our framework at the end of this work.

\subsection{Design Idea} \label{subsec:method_deisgn}

To meet the requirements for generating realistic network logs of smart grids during multi-stage cyberattacks, we propose a design that ensures realistic simulation, modularity, and scalability without compromise.

Each component in the simulation must mimic its physical counterpart's behavior. This includes modeling interactions with other elements in the simulation environment, especially data exchanges and remote control messages, to produce realistic data.

A modular structure allows flexible configuration of scenarios using reusable components. Containerization tools, such as Docker, support this modularity. Using Docker images for the smart grid actors enables easy instantiation of multiple components with the same logic but different parameters. Networking configurations ensure that only selected connections are enabled between containers, specifying maximum bandwidth and simulated latency to create realistic network logs, even on a single host. Users can define the topology in readable configuration files, adapting it as needed.

Orchestration tools help manage complex network topologies and prevent user errors. These tools can derive a whole simulation scenario from a single file, handling initialization and setup. Established orchestration tools also support scripts triggered after containers start or stop, facilitating the setup and extraction of network logs.


With these design choices, we chose a Docker-based approach leveraging containerization capabilities. Using a collection of pre-existing components for network communication and power simulation, we adopted the co-simulation environment proposed in \cite{cosim}.

This approach employs several tools and frameworks that align with our design decisions:

\begin{itemize}
    \item \textbf{rettij}: Utilizes Kubernetes orchestration, allowing the use of configuration files to define components and network connections in virtual smart grids via Docker images.
    \item \textbf{Mosaik and Panda Power}: Add power simulation capabilities, enabling realistic power grid simulations \cite{pandapower2018}.
\end{itemize}

Based on this co-simulation environment, we developed an attacker container that autonomously performs specified cyberattacks. This attacker integrates seamlessly into the co-simulation, generating network logs and power simulation results applicable to real-world scenarios.

By combining these components, our framework facilitates the generation of realistic, consistent, and high-quality data for analyzing and improving smart grid cybersecurity.

\section{Approach}\label{sec:approach}
\subsection{Co-Simulation Environment} \label{subsec:approach_cosim}

Our work utilizes the co-simulation software Mosaik, which groups individual simulators in a common context to build scenarios modularly \cite{mosaik}. Each simulator provides the functionality of real-world components, enabling the reconstruction of existing physical scenarios and the reuse of logic across varying simulator combinations.

To simulate the power grid and communication within a laboratory setup, we use Panda Power and other network communication simulators within Mosaik. Panda Power discretely simulates loads at each time step, while network communication simulators manage the interactions between components. Mosaik synchronizes these simulators, iterating through each simulation step and ensuring timely execution based on the current simulation time.

Mosaik connects modular simulators, synchronizing their communication to create a discrete environment. It iterates through each simulation step, checking if enough time has passed to execute the next step and then instructs the corresponding simulator to proceed.

Panda Power generates measurements for the virtual power grid, simulating components based on a predefined configuration file. This file describes power generators, transformers, buses, and power lines. The framework extracts detailed information about current usages and loads across all components \cite{pandapower2018}.

Terminal unit nodes handle IEC 104 communication, running as either RTUs or MTUs. They connect to a Mosaik server, translating measurements and commands to Panda Power to update the simulation. Terminal units can be configured to use predefined values or send specific commands at designated times.

\begin{figure}[htbp]
    \centerline{\includegraphics[width=\columnwidth]{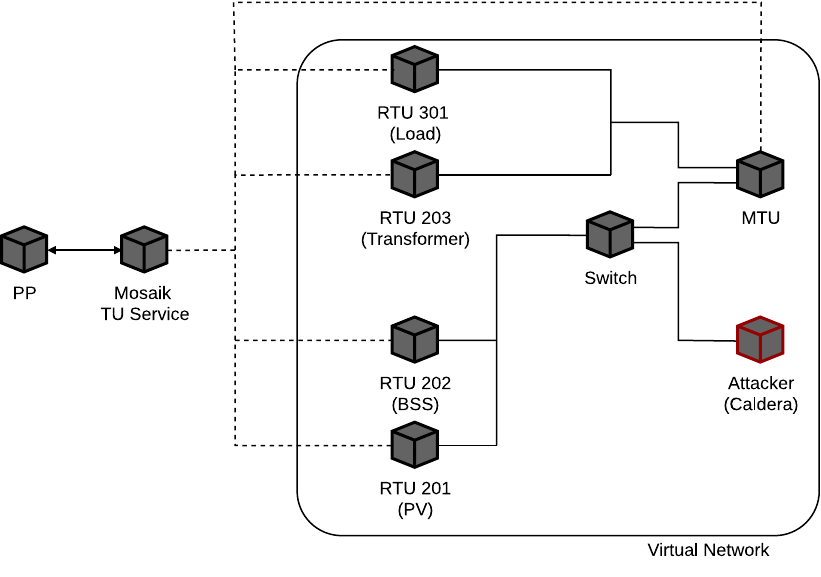}}
   \caption{Topology inside the co-simulation. All components are realized by individual containers. The virtual network emulates local laboratory communication, with only traffic through the switch considered in the network logs.}
   \label{fig:cosim-topology}
   \vspace{-0.5em}
\end{figure}


Rettij initializes virtual networks and starts containers, simplifying network connection definitions. The topology configuration specifies the components and their respective images (see Figure~\ref{fig:cosim-topology} as example). Rettij also handles virtual delays and bandwidth limitations, simulating realistic network conditions for larger grids.
With the help a virtual switch, that handles packet forwarding and is developed by the rettij team, we can create virtual networks between relevant components and capture all passing packets.
 The virtual network separates the communication of terminal units from Mosaik and Panda Power servers, keeping their messages out of the switch logs.
Managing networking between containers is a key challenge. The network must be segmented, with components communicating only with selected others. Rettij allows defining network topology in YAML format, modeling the LAN connections within the laboratory.


The rettij framework uses Kubernetes orchestration, defining components and network connections in virtual smart grids via Docker images. The combination of Mosaik, Panda Power, and rettij provides a robust environment for simulating and analyzing cyberattacks on smart grids.

By integrating modular attacks based on real-world cyberattacks, we ensure flexibility and reduce the need for specialized cybersecurity expertise. Our attacker container autonomously performs specified cyberattacks within the co-simulation, generating realistic and high-quality network logs and power simulation results.

\subsection{Attack Emulation} \label{subsec:approach_emulation}

To complete our scenarios, we need to integrate attacking logic into our simulation environment.
This section outlines our approach using MITRE Caldera™, a step-based attack emulation tool that aligns well with our co-simulation needs due to its structured and workflow-oriented nature.
Caldera employs a central command-and-control (C2) server with planning algorithms to manage attack steps, facilitating iterative execution of multi-stage attacks within the co-simulation.

Several alternatives exist for attack emulation, namely Atomic Red Team, Splunk and MulVal. But they do not focus on Linux systems for their code, are closed-source or only report assessments according to vulnerability reports.


Caldera on the other hand includes planning capabilities for autonomous execution, high extensibility due to plugins, a modular configuration and is actively developed as an open-source project.
The smallest component the attack execution relies on, are the so-called abilities.
These are YAML-defined scripts and commands that implement ATT\&CK techniques. Each ability is identified by a unique ID and includes executors specifying the platform and command to be run, such as Shell for Linux. Abilities may include payloads downloaded as needed and cleanup commands to minimize traces on victim machines. An example of how these might look is given in \autoref{fig:ability}.

\begin{figure}
\begin{scriptsize}
\begin{verbatim}
name: Capture Network Traffic (TCPDump with Scapy)
id: 1b27e1f8-af08-47eb-b3dc-100c1d697413
  platform: linux
    command: /bin/python tcpdump.py -t 150
    payloads: [tcpdump.py]
    cleanup: [/bin/rm tcpdump.py]
\end{verbatim}
\end{scriptsize}
\caption{Caldera ability for running a Python script.}
\label{fig:ability}
\vspace{-.5cm}
\end{figure}

Adversaries are constructs combining multiple abilities to form multi-stage attack plans. Each adversary definition ensures that abilities are executed in sequence, with dependencies checked for each step. For example, an adversary performing a \ac{DoS} attack might include abilities for network scanning, host enumeration, and attack execution.

Fact sources store and manage data used across agents and abilities, allowing for dynamic execution based on collected or pre-defined information. Facts are labeled values, such as IP addresses or credentials, that can trigger abilities. Rules filter which facts are used, ensuring targeted execution.


Operations combine adversaries and fact sources, specifying execution details such as agent groups, step intervals, and execution mode (automatic or manual). Operations control the overall attack flow, leveraging modular definitions to reuse attack logic and necessary information.

To execute an attack, we follow these steps:

\begin{enumerate}
    \item \textbf{Start the C2 Server}: The C2 server orchestrates components, provides payload access, and listens for agent connections. It runs locally without internet access, maintaining all necessary attack information.
    \item \textbf{Connect an Agent}: An agent, a client executable on a compromised machine, connects to the C2 server. It communicates via HTTP packets, providing minimal obfuscation. Agents are categorized by group identifiers for targeted attack execution.
    \item \textbf{Execute an Operation}: We select the agent group and adversary for execution. The operation uses Caldera’s planning module to determine and send instructions to agents based on available facts and adversary requirements. Agents execute commands, collect outputs, and send results back to the server for parsing and further execution.
    \item \textbf{Cleanup}: To minimize traces, agents execute cleanup commands either remotely by the server or as part of adversary completion.
\end{enumerate}

This structured approach ensures detailed control and parallel execution across multiple agents, enhancing the robustness of our attack emulation within the simulated environment.

\vspace{-0.5em}

\subsection{Generative Methods}

We tested three state-of-the-art open-source \acp{LLM} pre-trained on large code bases and focused on code generation, given our hardware constraints (NVIDIA GeForce RTX 3060 Ti with 8 GB VRAM). 
Code Llama \cite{codellama} (13 billion parameters), DeepSeek Coder \cite{guo2024deepseek} (6.7 billion parameters), and Starcoder 2 \cite{lozhkov2024starcoder} (7 billion parameters). 


We tested the \acp{LLM}’ understanding and generation capabilities through four tasks, including the analysis and generation of abilities and adversaries.

We transformed all local definitions of Caldera's abilities and adversaries into an embedding, which the \acp{LLM} could use as reference during their inference. 
This process, \ac{RAG}, was handled by PrivateGPT \cite{Martinez_Toro_PrivateGPT_2023}, allowing us to ingest all local YAML files automatically.

\section{Results}

All results of the co-simulation were run in the scenario depicted in \autoref{fig:cosim-topology} with the same attacker and a set delay for running the final step of the attack after ten minutes.

Inspired by the \ac{CIA} triad we ran three different attacks in addition to the default scenario without any attack going on.
For testing the confidentiality we list all accessible files on the infected machine and sent the results back to the attacker, while the integrity of the system was compromised by manipulating the control commands sent by the \ac{MTU}.
Finally disabling the network connections of the infected hosts assessed the impact on availability.
In the following we present excerpts from our results.

\subsection{Framework's Applicability}
\begin{figure}[htbp]
    \centerline{\includegraphics[width=\columnwidth]{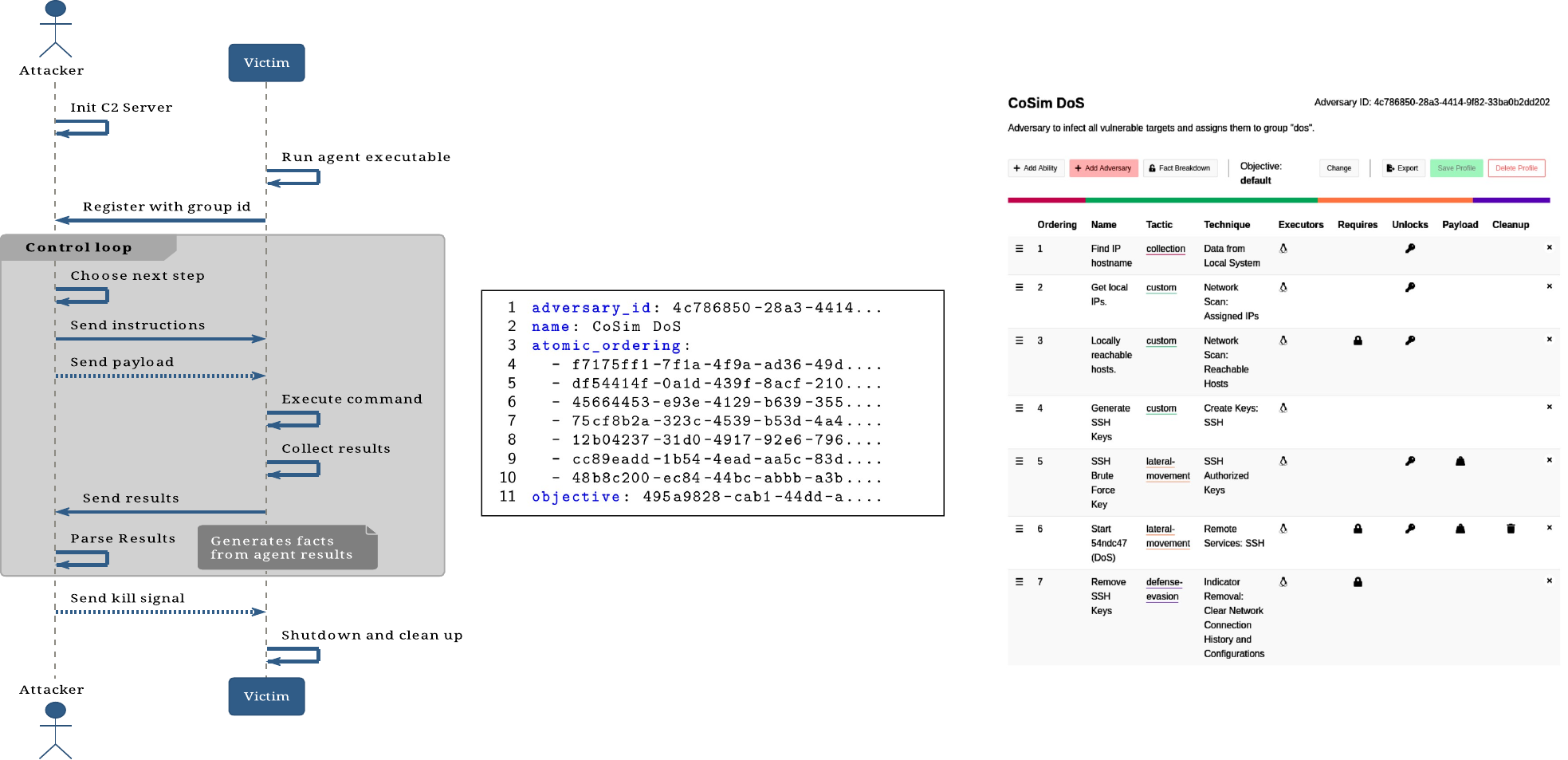}}
    \caption{Control sequence between an attacker using Caldera and a compromised victim.}
    \label{fig:framework_applicability}
    \vspace{-1em}
\end{figure}

The framework uses YAML files to define simulation parameters, including network topology, simulators, and connections. Key elements include controller and proxy modules for different backends, and global parameters like runtime and step size for synchronization. The network topology specifies IP addresses, bandwidth, and delay. The co-simulation deploys components, including SCADA devices, networking hardware, and power grid components. The developed approach integrates with co-simulation environments using a structured workflow and central C2 server (cf. Figure~\ref{fig:framework_applicability}). This server autonomously executes multi-stage cyberattacks, ideal for Mosaik simulations. Components include abilities (predefined scripts), adversaries (execution plans), fact sources (data stores), and operations (execution flow).

In a simulated environment, the approach can execute DoS attacks. Vulnerable terminal units mimic real-world issues like default SSH passwords. An attacker node initializes the C2 server and connects an agent on a compromised machine. The operation starts with a brute-force SSH attack and disables network interfaces. Attacker actions include NMAP scans, Telnet connections, SCP file transfers, netcat use, and shell commands. Configurations test system performance, standard operation, power manipulation, and scenarios causing power spikes, vRTU slowdowns, or shutdowns. Other setups involve Telnet data exfiltration and reconnaissance via NMAP and Telnet logins. The simulation generates outputs, capturing network traffic, logging actions, and creating reports for analysis.

In cyber-physical labs, the approach's lightweight scripts and modular components adapt to various attack scenarios. The C2 server commands infected agents, executing abilities and returning results. This data helps evaluate the effectiveness of countermeasures and understand cyberattack impacts on smart grids.

\subsection{Power Simulation}

\begin{figure}[htbp]
    \centerline{\includegraphics[width=.7\columnwidth]{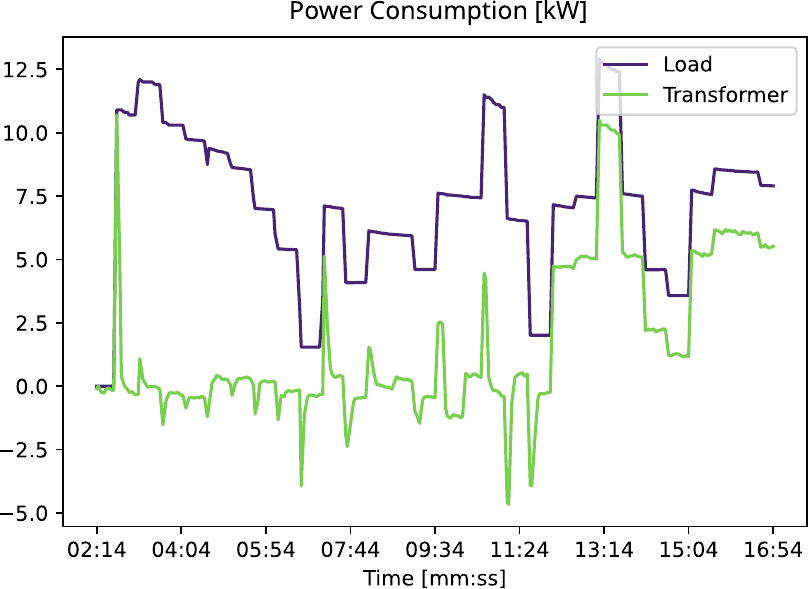}}
    \caption{Measured power curve at substation of the \acs{DoS} scenario in the laboratory.}
    \label{fig:power_dos_lab}
    \vspace{-1em}
\end{figure}

\begin{figure}[htbp]
    \centerline{\includegraphics[width=.7\columnwidth]{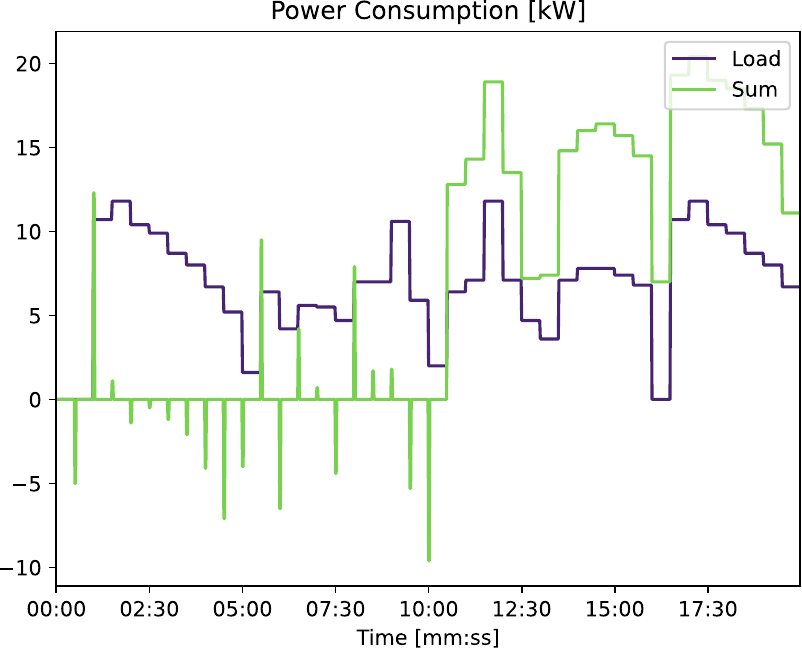}}
    \caption{Measured power curve at substation of the \ac{DoS} scenario run in the co-simulation.}
    \label{fig:power_dos_cosim}
    \vspace{-0.5em}
\end{figure}

Looking at the measured power levels via the three-phase power measurement device of the laboratory (\autoref{fig:power_dos_lab}), we see the balancing of the components in a self-consumption optimization operation scenario.
The initial ten minutes of the \ac{DoS} scenario are almost identical to running the laboratory without any attacker in it.
But then the communication between the \ac{MTU} and \ac{RTU} is disrupted by the attack and the regulatory commands do not reach the corresponding \ac{DER}.
Thus the imbalances occur and the grid sum now mirrors the load induces to the power grid.

Inside the co-simulation we observe the same effect of imbalances after the attack is executed (cf. Figure~\ref{fig:power_dos_cosim}).
Noteworthy is the exact timing of when the impact happens, which is contrasted by the timing in the laboratory.
This is due to the manual setup that is required when running the physical components, which can easily be automated inside the co-simulation.


\subsection{Timely Spread}

\begin{figure}
\begin{subfigure}{.5\columnwidth}
    \centerline{\includegraphics[width=\textwidth]{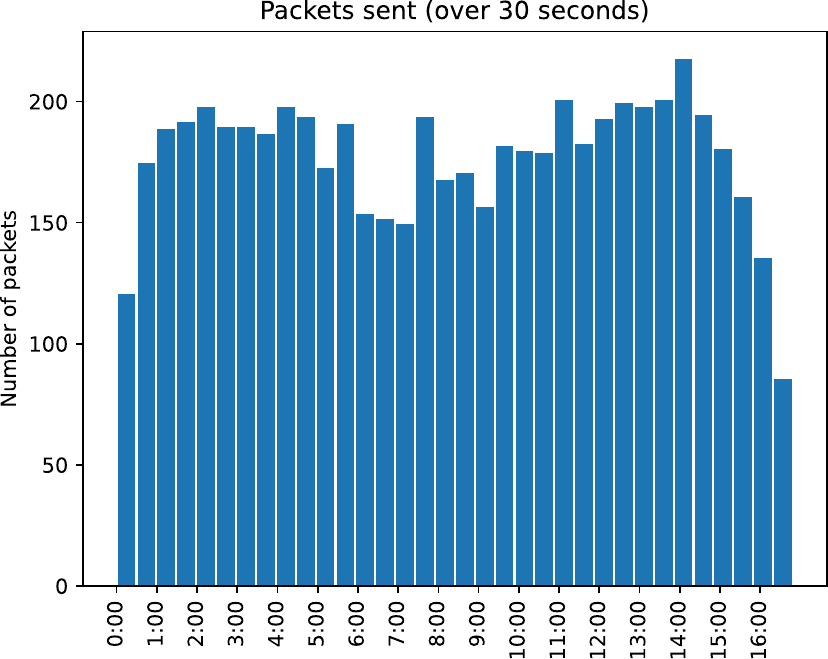}}
    \caption{Laboratory.}
\end{subfigure}%
\begin{subfigure}{.5\columnwidth}
    \centerline{\includegraphics[width=\textwidth]{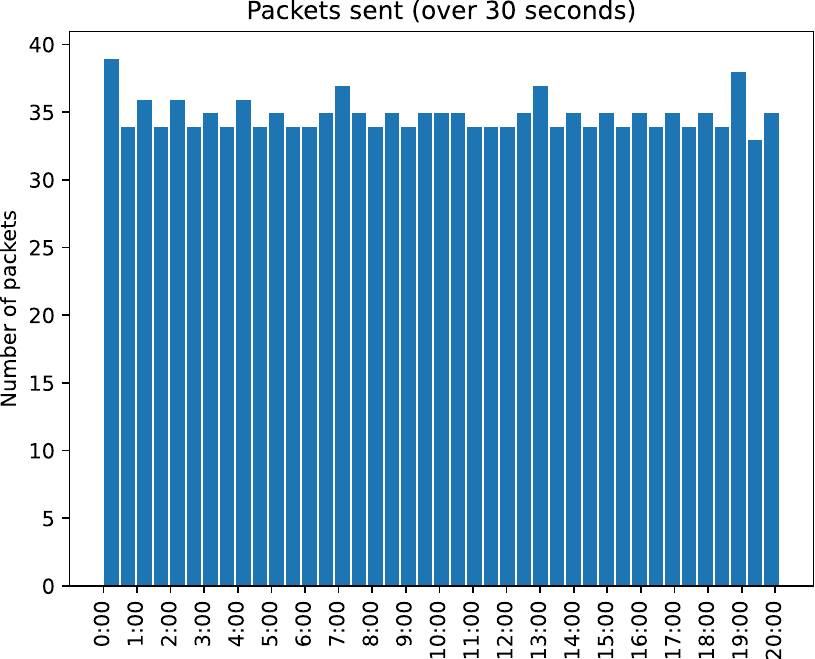}}
    \caption{Co-simulation}
\end{subfigure}
    \caption{Packet count over time in normal scenarios.}
    \label{fig:packet_count}
\vspace{-1em}
\end{figure}

As a next metric we inspected the number of packets sent over time. In \autoref{fig:packet_count} we see how the amount of network traffic fluctuates over time inside the laboratory, while the co-simulation produces a more stable and homogeneous amount of network packets during the scenario.

One of our requirements is the consistency of the data, which is hard to ensure inside physical setups due to the warm start of the environment. This means that organisational communications happen before the actual scenario is run, rendering the recorded network traffic incomplete. By fully instantiating the network during the co-simulation, we can collect all of this traffic and reproduce even the organisational communication during each scenario.

\FloatBarrier
\subsection{Connections}

Next in line is the connection overview. Here we show how much traffic occurs between the individual hosts.
As one key result we see how much data the connections for controlling the terminal units make up.
This communication is only required because the \acp{VRTU} need to be started by hand to correctly time their execution.

\begin{figure}[htbp]
    \centerline{\includegraphics[width=.8\columnwidth]{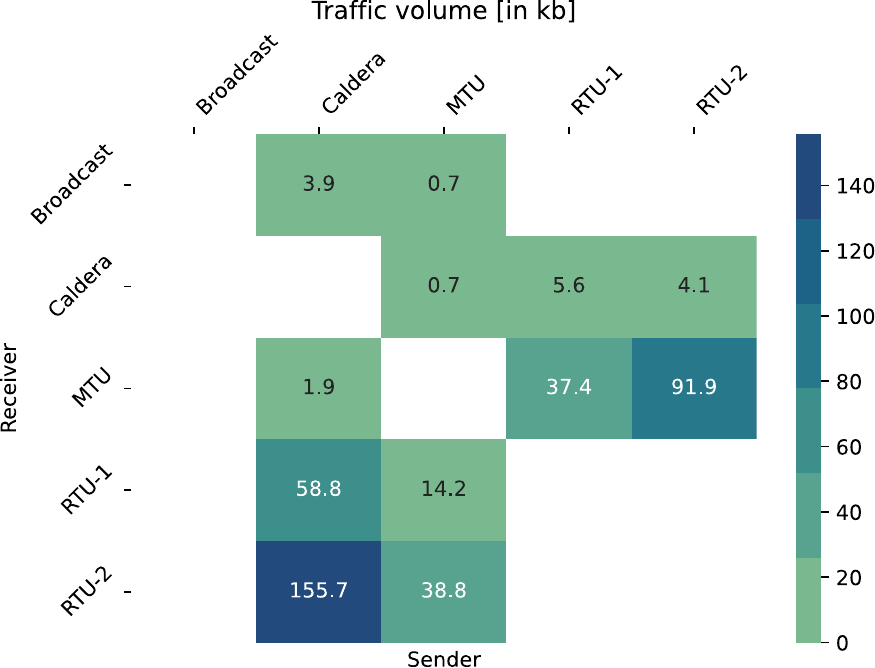}}
    \caption{Network connections in the laboratory.}
    \label{fig:connections_normal_lab}
    \vspace{-1em}
\end{figure}

\begin{figure}[htbp]
    \centerline{\includegraphics[width=.8\columnwidth]{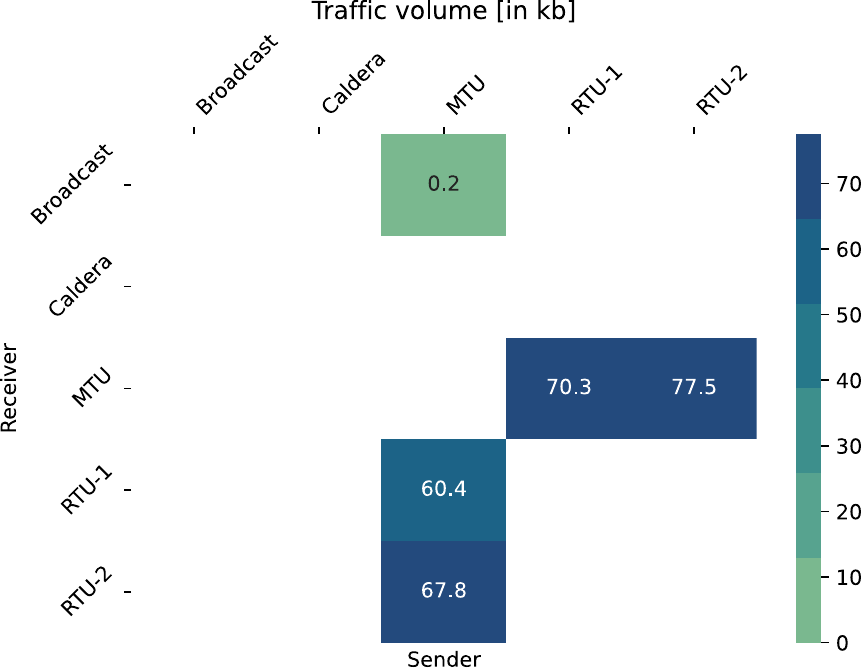}}
    \caption{Network connections in the co-simulation.}
    \label{fig:connections_normal_cosim}
    \vspace{-0.5em}
\end{figure}

In \autoref{fig:connections_normal_lab} we see the default communication without any attack performed.
Here the \textit{Caldera} entry refers to the host, that connects to the terminal units to run the required scripts.

When we compare these results to those of the co-simulation in \autoref{fig:connections_normal_cosim}, we clearly see how the network log only contains the communication between the \ac{MTU} and the \acp{RTU} and nothing else.
Working with clean data allows the data to make sure even slightest irregularities due to the attackers behaviour can be accounted for, while the laboratory's heterogeneous equipment is prone to potential variations.

\subsection{Protocol Distribution}

The impacts of the attacks can also be seen in the composition of different protocols that are contained in the network logs. In \autoref{tab:protocols} we listed the percentages that the most frequent protocols contributed to the overall packet count.

\begin{table}
    \centering
    \caption{Protocol percentages and absolute counts of network traffic for different scenarios (normal, manipulation, and DoS), with total packet numbers below each scenario name.}    
    \begin{tabular}{l|rrr|rrr}
    \toprule \\
         & \multicolumn{3}{c}{Laboratory}  & \multicolumn{3}{c}{Co-Simulation} \\
                    & Normal     & Manip.     & DoS     & Normal     & Manip.     & DoS \\
         Protocol   & (1509)     & (3,577)     & (2,205)     &  (1,431)    & (2,588)     & (1,792) \\
        \midrule \\
         IEC 104    & 0.34 & 0.15 & 0.13 & 0.67  & 0.39 & 0.28 \\
         SSH        & 0.54 & 0.77 & 0.79 & 0.00  & 0.36 & 0.46 \\
         ARP        & 0.01 & 0.04 & 0.07 & 0.32  & 0.24 & 0.24 \\
         LLC        & 0.07 & 0.04 & 0.00 & 0.00  & 0.00 & 0.00  \\
         Other      & 0.04 & 0.01 & 0.01 & 0.02  & 0.02 & 0.03 \\
        \midrule \\
         IEC 104    & 510   & 542     &  280  & 956   & 1008  & 500 \\
         SSH        & 811   & 2743    & 1744  & 0     & 926   & 818 \\
         ARP        & 20    & 144     &  164  & 452   & 614   & 429 \\
         LLC        & 113   & 129     &    0  & 0     & 0     & 0   \\
         Other      & 55    & 19      &   17  & 23    & 40    & 45  \\
         \bottomrule
    \end{tabular}
    \label{tab:protocols}
\vspace{-1em}
\end{table}

In the laboratory we captured a lot of \ac{SSH} packets originating from the host that had to start the processes manually on the \acp{VRTU}. These packets obfuscate the the traffic, that is introduced by the attacker and its \ac{SSH} brute-force ability. In the co-simulation we do not have to rely on manual connections for re-/starting the processes on any component, which reduces the noise inside the exported network log.
Another example is the \ac{ARP} that is only partially captured in the laboratory, while we see all packets in the log of the co-simulation.
We also see protocols like the \ac{LLC} that are introduced by the hardware switch, which only occurred inconsistently during the test in the laboratory and are an example of how unwanted packets can pollute the captured network logs. Even though filtering is a valid option, this involves granular post-processing of the collected data. 
The \ac{DoS} attack nearly halves IEC 104 packet counts in both environments and reduces \ac{SSH} packet amounts.

\FloatBarrier
\subsection{Further Findings}

The generation of data in the laboratory proved to be more volatile due to manually starting the communication process on the involved components, which poses a big challenge for reproducibility.
Similarly we observed more noise due to specific hardware like the switches and network taps, that are required for recording network traffic inside the local network.

Furthermore, the fact that the \ac{VRTU} in the physical laboratory did not expose additional network interfaces for separately running the SSH connections to start their internal workings, introduced a lot of network traffic, that would not occur on productive \ac{RTU} components.

To conclude the experience from setting up the test scenario inside a physical laboratory, we can say that minimizing the noise was difficult and in the end not completely possible due to the available hardware and communication channels. In general the modular configuration of the co-simulation allows for less restrictive setups and also reduces potential interference with unwanted processes.

Also worth mentioning is the fact, that our autonomous agent, that was developed inside the co-simulation, could immediately run most of the attack steps without any adaption to its configuration. Only the binary that is sent to the target machine needed to be recompiled for the ARM architecture of the \ac{VRTU}. Besides that, the whole attack scenario could just be run as is.

Finally the results from querying the \acp{LLM} did not yield immediately usable code fragments. Instead the generated text contained partially usable code, that still needed to be adapted to make use of Caldera's facts mechanism. Thus the usage of these models might be helpful during the development, they are no source for reliable code, that can be embedded directly into a running attacker.

\begin{figure}[htbp]
\begin{subfigure}{.5\columnwidth}
    \centerline{\includegraphics[width=\textwidth]{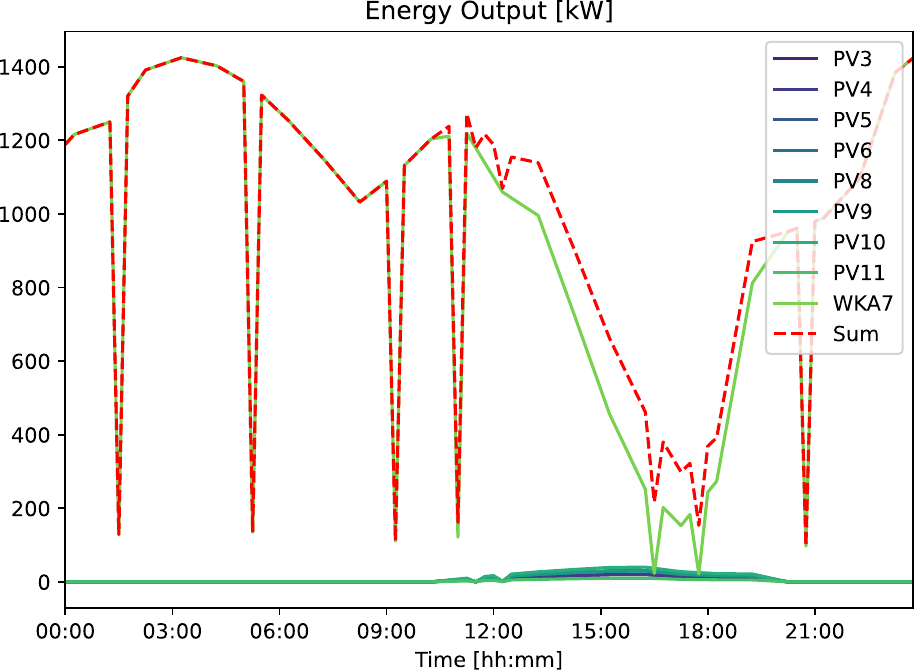}}
    \caption{Normal}
\end{subfigure}%
\begin{subfigure}{.5\columnwidth}
    \centerline{\includegraphics[width=\textwidth]{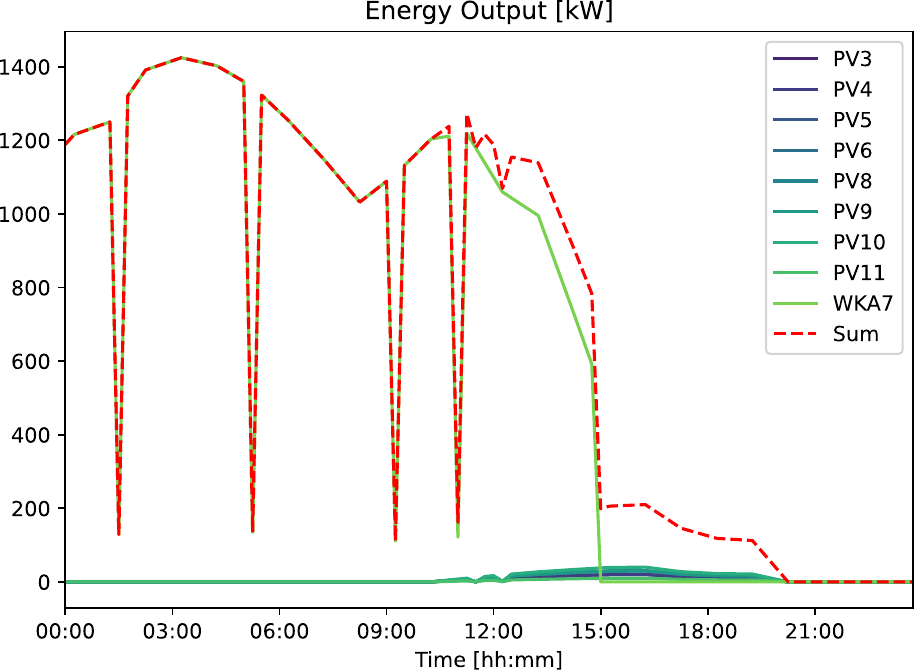}}
    \caption{Manipulation}
\end{subfigure}
    \caption{Power output in the CIGRE MV\cite{cigremv} scenario.}
    \label{fig:cigre}
\vspace{-0.5em}
\end{figure}

Finally we also successfully ran our framework on the topology suggested by the CIGRE Task Force\cite{cigremv}. Again the attacker was able to adapt to the more complex network without further configurations (see \autoref{fig:cigre}).
Only the target values for the manipulation had to be set.
This indicates the potential flexibility of our framework, but requires further research.
\vspace{-1em}

\section{Conclusion}

With this work we demonstrated how the overall data scarcity for attack data on smart grids can be tackled.
We introduced a framework, that allows for configurable data generation and compared the impacts on the physical and simulated smart grids, showing that even though the data is not identical, the impacts the attacks have can be well reproduced.
In general the virtually generated data contains less noise due to precise definitions of what happens in the network.

Our proposed methods allowed us to not only develop cyberattacks that could be deployed in the physical laboratory with almost no additional work, but also generated data that achieves higher consistency between different runs. The easily adaptable setup of the virtual environment enabled a fast reproduction of the complex physical setup and thus increased the overall development efficiency.

In the future we hope to see more complex cyberattacks on larger grid topologies.
By allowing for quick prototyping and testing of attack methods we also see a lot of potential for more in depth analysis of overall cybersecurity of smart grids without the limiations of openly available datasets.

\section*{Acknowledgment}
\begin{minipage}{0.65\columnwidth}%
Received funding from the BMBF under project no. 03SF0694A (Beautiful).
\end{minipage}
\hspace{0.02\columnwidth}
\begin{minipage}{0.23\columnwidth}%
	\includegraphics[width=\textwidth]{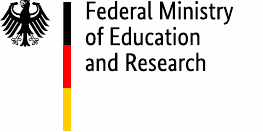}
\end{minipage}
\vspace{-0.5em}

\bibliographystyle{IEEEtran}
\bibliography{fullpaper}

\begin{thebibliography}{10}
\providecommand{\url}[1]{#1}
\csname url@samestyle\endcsname
\providecommand{\newblock}{\relax}
\providecommand{\bibinfo}[2]{#2}
\providecommand{\BIBentrySTDinterwordspacing}{\spaceskip=0pt\relax}
\providecommand{\BIBentryALTinterwordstretchfactor}{4}
\providecommand{\BIBentryALTinterwordspacing}{\spaceskip=\fontdimen2\font plus
\BIBentryALTinterwordstretchfactor\fontdimen3\font minus \fontdimen4\font\relax}
\providecommand{\BIBforeignlanguage}[2]{{%
\expandafter\ifx\csname l@#1\endcsname\relax
\typeout{** WARNING: IEEEtran.bst: No hyphenation pattern has been}%
\typeout{** loaded for the language `#1'. Using the pattern for}%
\typeout{** the default language instead.}%
\else
\language=\csname l@#1\endcsname
\fi
#2}}
\providecommand{\BIBdecl}{\relax}
\BIBdecl

\bibitem{kenyon2020public}
A.~Kenyon \emph{et~al.}, ``Are public intrusion datasets fit for purpose characterising the state of the art in intrusion event datasets,'' \emph{Computers \& Security}, 2020.

\bibitem{le2019smart}
T.~D. Le \emph{et~al.}, ``Smart grid co-simulation tools: Review and cybersecurity case study,'' in \emph{icSmartGrid}, 2019.

\bibitem{plan2defend}
T.~Choi \emph{et~al.}, ``Plan2defend: Ai planning for cybersecurity in smart grids,'' in \emph{ISGT Asia}, 2021.

\bibitem{vyas2023automated}
S.~Vyas \emph{et~al.}, ``Professor pete burnap. automated cyber defence: A review,'' \emph{arXiv preprint arXiv:2303.04926}, 2023.

\bibitem{williams1994purdue}
T.~J. Williams, ``The purdue enterprise reference architecture,'' \emph{Computers in industry}, 1994.

\bibitem{formosa2021robot}
P.~Formosa, ``Robot autonomy vs. human autonomy: social robots, artificial intelligence (ai), and the nature of autonomy,'' \emph{Minds and Machines}, 2021.

\bibitem{EU-AIA}
\BIBentryALTinterwordspacing
{Council of European Union}, ``Artificial intelligence act,'' 2024. [Online]. Available: \url{https://eur-lex.europa.eu/legal-content/EN/TXT/?uri=CELEX:52021PC0206}
\BIBentrySTDinterwordspacing

\bibitem{li2012}
X.~Li \emph{et~al.}, ``Securing smart grid: cyber attacks, countermeasures, and challenges,'' \emph{IEEE Communications Magazine}, 2012.

\bibitem{ahmadian2016causal}
A.~Ahmadian~Ramaki \emph{et~al.}, ``Causal knowledge analysis for detecting and modeling multi-step attacks,'' \emph{secur. commun. netw.}, 2016.

\bibitem{cosim}
D.~van~der Velde \emph{et~al.}, ``Towards a scalable and flexible smart grid co-simulation environment to investigate communication infrastructures for resilient distribution grid operation,'' in \emph{SEST}, 2021.

\bibitem{pandapower2018}
L.~Thurner \emph{et~al.}, ``pandapower — an open-source python tool for convenient modeling, analysis, and optimization of electric power systems,'' \emph{IEEE Transactions on Power Systems}, 2018.

\bibitem{mosaik}
A.~Ofenloch \emph{et~al.}, ``Mosaik 3.0: Combining time-stepped and discrete event simulation,'' in \emph{OSMSES}, 2022.

\bibitem{codellama}
B.~Roziere \emph{et~al.}, ``Code llama: Open foundation models for code,'' \emph{arXiv:2308.12950}, 2023.

\bibitem{guo2024deepseek}
D.~Guo \emph{et~al.}, ``Deepseek-coder: When the large language model meets programming--the rise of code intelligence,'' \emph{arXiv:2401.14196}, 2024.

\bibitem{lozhkov2024starcoder}
A.~Lozhkov \emph{et~al.}, ``Starcoder 2 and the stack v2: The next generation,'' \emph{arXiv:2402.19173}, 2024.

\bibitem{Martinez_Toro_PrivateGPT_2023}
\BIBentryALTinterwordspacing
I.~Martínez~Toro \emph{et~al.}, ``{PrivateGPT},'' 2023. [Online]. Available: \url{https://github.com/imartinez/privateGPT}
\BIBentrySTDinterwordspacing

\bibitem{cigremv}
K.~Rudion \emph{et~al.}, ``Design of benchmark of medium voltage distribution network for investigation of dg integration,'' in \emph{IEEE PESGM}, 2006.

\end{thebibliography}

\end{document}